\documentclass[journal]{IEEEtran}
\usepackage{cite}

\ifCLASSINFOpdf
   \usepackage[pdftex]{graphicx}
\else
\fi
\usepackage{fixltx2e}

\usepackage{stfloats}
\begin{document}
%
\title{Spincaloritronic measurements: a round robin comparison of the longitudinal spin Seebeck effect}
%
%
%

\author{Alessandro~Sola,~
		Vittorio~Basso,~
        Michaela~Kuepferling,~
        Massimo~Pasquale,
        Daniel~Meier,~
        G\"{u}nter~Reiss,~
        Timo~Kuschel,~
        Takashi~Kikkawa,~
        Ken-ichi~Uchida,
        Eiji~Saitoh,~
        Hyungyu~Jin,~
        Steve~Boona,~
        Sarah~Watzman,~
        Joseph~Heremans,
        Matthias~B.~Jungfleisch,~
        Wei~Zhang,~
        John~E.~Pearson,
        Axel~Hoffmann,~
        Hans~W.~Schumacher
\thanks{A. Sola, V. Basso, M. Kuepferling and M. Pasquale are with the Istituto Nazionale di Ricerca Metrologica, 10135 Turin, Italy (e mail: a.sola@inrim.it)}
\thanks{D. Meier, G. Reiss and T. Kuschel are with the Center for Spinelectronic Materials and Devices, Bielefeld University, 33615 Bielefeld, Germany.}
\thanks{T. Kikkawa, K. Uchida and E. Saitoh are with the Institute for Materials Research, Tohoku University, Sendai 980-8577, Japan.}
\thanks{K. Uchida is with the National Institute for Materials Science, Tsukuba 305-0047, Japan.}
\thanks{H. Jin, S. Boona, S. Watzman and J. Heremans are with the Department of Mechanical and Aerospace Engineering, The Ohio State University, Columbus, Ohio 43210, USA.}
\thanks{M. B. Jungfleisch is with the Department of Physics and Astronomy, University of Delaware, Newark, DE 19716, USA}
\thanks{W. Zhang is with the Department of Physics, Oakland University, Rochester MI 48309, USA}
\thanks{J. E. Pearson and A. Hoffmann are with the Materials Science Division, Argonne National Laboratory, Argonne, Illinois 60439, USA.}
\thanks{H. W. Schumacher is with the Physikalisch-Technische Bundesanstalt, Bundesallee 100, 38116 Braunschweig, Germany.}
\thanks{This work has been submitted to the IEEE for possible publication. Copyright may be transferred without notice, after which this version may no longer be accessible.}}

\maketitle

\begin{abstract}
The rising field of spin caloritronics focuses on the interactions between spin and heat currents in a magnetic material; the observation of the spin Seebeck effect opened the route to this branch of research. This paper reports the results of a round robin test performed by five partners on a single device highlighting the reproducibility problems related to the measurements of the spin Seebeck coefficient, the quantity that describes the strength of the spin Seebeck effect. This work stimulated the search for more reproducible measurement methods through the analysis of the systematic effects.
\end{abstract}


%
\IEEEpeerreviewmaketitle

\section{Introduction}

%
%
%

\IEEEPARstart{T}{he} concept of spintronics gained interest in the last 30 years, since the observations of spin-dependent electron transport phenomena in solids\cite{johnson1988coupling,wolf2001spintronics,fert2008nobel}.
From the viewpoint of technology, the possibility of handling the spins in electronic devices is revolutionary for two reasons: first, adding the spin degree of freedom to the charge allows a new type of data processing. Taking into account the spin of the electron, in fact, can lead to new phenomena as for example the giant magnetoresistance in spin valves\cite{chappert2007emergence}. Second, novel electronic devices may be based only on the spin instead of on the charge of the electron. This is possible in magnetic insulators, where spin currents carried by spin waves (or magnons i.e. the quanta of spin waves) exist independently of charge currents\cite{serga2010yig,chumak2015magnon}. There are potentially huge advantages in terms of speed and cooling of devices with the development of novel electronics (spintronics, indeed) which are based on the use of spins instead of electric charges.
The idea of generating spin currents in solids was demonstrated by the study of phenomena like the spin pumping\cite{tserkovnyak2002spin,heinrich2011spin,mosendz2010quantifying,sandweg2011spin} and the spin Hall effect\cite{hirsch1999spin,kato2004observation}, by means of microwave excitation and by an electric current flowing in an adjacent layer of a high spin-orbit coupling material (e.g. platinum).

More recently since 2008, the research group led by E. Saitoh at Tohoku University followed by others investigated the generation of a spin current in a magnetic material as a consequence of a temperature gradient \cite{Uchida0,Uchida1,Jaworski,Uchida2,Weiler,Meier0,huang2011intrinsic,avery2012observation}; this phenomenon is called spin Seebeck effect (SSE) as reference to the spin counterpart of the Seebeck effect. It is possible to detect electrically the spin current generated by the SSE by means of the inverse spin Hall effect (ISHE) \cite{Saitoh}; this rises in a high spin-orbit coupling material, deposited on the magnetic material that produces a spin current.
Since the first investigations on the SSE, different configurations in terms of geometry and materials have been reported in the literature. For what concerns the ISHE layer, the contribution to the ISHE voltage depends on its thickness\cite{castel2012platinum} and on the atomic number of the metal\cite{wang2014scaling}, according to the spin-charge conversion efficiency, that is the spin Hall angle.
The thickness of the active layer for the SSE (i.e. the magnetic material) has been taken into account in order to investigate the characteristic length-scale for the phenomenon\cite{kehlberger2015length,yamamoto2017evaluation}. Also the chemical composition of the magnetic layer has been the subject of a wide research\cite{uchida2010longitudinal,uchida2013longitudinal,siegel2014robust,Meier0}.

Possible applications of the widely investigated SSE materials are sensors\cite{kirihara2016flexible,liao2017spin,siegel2015spin}, spin-analogues of thermoelectric generators\cite{kirihara2012spin,uchida2016thermoelectric,uchida2014longitudinal,cahaya2015spin} and devices designed in view of the development of spintronic circuits, such as spin batteries\cite{yu2017spin}. For what concerns the spin-thermoelectric effects, if we make a parallel with the usual thermoelectric materials\cite{goldsmid2001} where $ZT = \left( \varepsilon^{2} \sigma\right)/\left( kT\right) $, we expect a good figure of merit for materials with poor thermal conductivity $k$ and large spin conductivity $\sigma$ as are found in insulating ferrimagnets, like YIG and ferrites. Spin thermoelectric devices are being designed and fabricated as thin films\cite{kirihara2012spin,uchida2016thermoelectric,uchida2014longitudinal,cahaya2015spin} and multilayers\cite{ramos2015unconventional,ramos2016thermoelectric} and have room for improvement with a detailed quantitative knowledge of the intrinsic characteristics. In this wide scenario of experiments and possible applications, a reproducible quantitative determination of the SSE effect is crucial.

\section{Measurement of the spin Seebeck effect}
Due to the novelty of the SSE, an exhaustive interpretation of the phenomenon was initially lacking, even from the experimental viewpoint.
As a first step in the measurement of the SSE, it is necessary to take into account the magnetic origin of the effect in order to separate its contribution from the spurious component of ordinary Seebeck effect. The ohmic contacts of a SSE device are usually made up of interfaces between different metals such as platinum, gold or silver. Being these interfaces subjected to a temperature gradient, they may cause the rising of an ordinary Seebeck effect contribution that is summed to the SSE one. This contribution is negligible except when the geometry of the two electrical contacts exhibits some asymmetries. In order to cancel the ordinary Seebeck effect component, it is necessary to exploit the odd parity of the first one, opposed to the even parity of the second one as function of the applied magnetic field. In this way, the SSE component is measured as the half difference of the Seebeck observables (SSE and ordinary Seebeck effect) at two opposite values of magnetic field, which has to be strong enough to saturate the material, in case this exhibits hysteresis. By using this procedure, the ordinary Seebeck effect which does not depend on the magnetic field can be treated as an additive constant and cancelled out.
Other important challenges in the experimental research on the SSE are the study of some possible contributions of other magneto-thermoelectric effects, like for example the anomalous Nernst effect \cite{kikkawa2013longitudinal,kikkawa2013separation,schmid2013transverse,meier2013influence,huang2011intrinsic,avery2012observation,bougiatioti2017quantitative,shestakov2015dependence,reimer2017quantitative} and the geometry of the SSE (longitudinal vs. transverse configuration)\cite{Uchida2,Meier}.

At present, the most studied configuration is the longitudinal spin Seebeck effect (LSSE); this rises in a bilayer system formed by an in-plane magnetized layer, typically a ferrimagnetic yttrium iron garnet (YIG), covered by a Pt thin film for the ISHE detection. In this configuration, the temperature gradient is applied out of plane in order to inject a spin current in the top Pt film.
The quantitative determination of the LSSE is represented by the LSSE coefficient $S_{\textrm{LSSE}}$, defined by the expression

\begin{eqnarray}
S_{\textrm{LSSE}} = \left( \frac{V_{\textrm{ISHE}}}{L}\right)/\left( \frac{\Delta T}{L_{\textrm{x}}}\right).
\end{eqnarray}

\begin{figure}[!ht]
	\centering
	\includegraphics[width=3.4in]{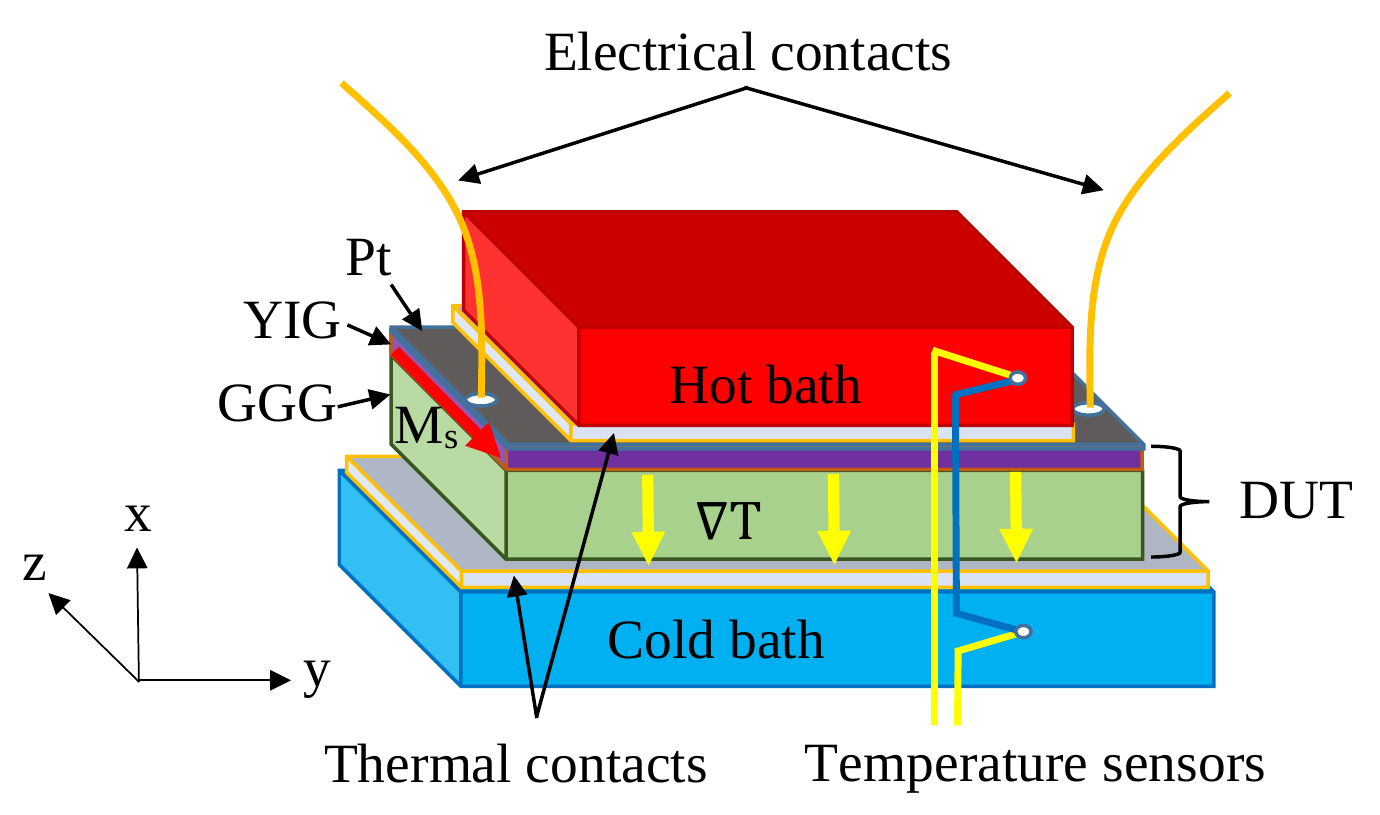}
	\caption{Scheme of the LSSE experiment: the in-plane magnetized LSSE sample is clamped between the two thermal baths whose temperature is probed by the sensors (e.g. thermocouples). The blue and the yellow wires of the temperature sensors represent the two materials of the thermocouple. Between the thermal baths and the sample a thermal conductor and a thermal contact are present. The electrical contacts are set on the top edges of the Pt and they are probing the voltage drop along the $y$ axes. The magnetization $M$ is directed along the $z$ axes, and the temperature gradient is set out-of-plane.}
	\label{fig_1}
\end{figure}

\noindent The voltage $V_{\textrm{ISHE}}$ represents the electrical observable and is proportional to the charge accumulation along the Pt film of length $L$, as consequence of the ISHE. The thermal observable is the temperature difference $\Delta T$ across $L_{\textrm{x}}$, that is the thickness of the sample, having considered the thermal conductivity of the active layer (YIG) equal to the one of the substrate.
A scheme of the measurement configuration for the LSSE is reported in Fig. \ref{fig_1}. 
\noindent Once defined the LSSE coefficient $S_{\textrm{LSSE}}$, its quantitative determination consists of three measurements: the electric one ($V_{\textrm{ISHE}}$), the thermal one ($\Delta T$) and the measurement of magnetic field.
The latter is not essential for the quantitative determination of the coefficient $S_{\textrm{LSSE}}$, since the magnetic material has to be treated as a single domain, i.e. at magnetic saturation $\pm M_{\textrm{s}}$, as previously discussed, and the $V_{\textrm{ISHE}}$ as function of the magnetic field characteristics retraces the magnetization loop of the active layer surface. The electric observable depends on the ohmic contact that is fabricated on the Pt film, at the two opposite edges along the $y$ direction.
For the measurement of the temperature difference $\Delta T$, some techniques are reported in the literature, including measurements of heat currents\cite{Sola1,Sola2,bougiatioti2017quantitative,prakash2018evidence} and measurements of the resistance of two Pt layers deposited on the two surfaces of the magnetic layer\cite{uchida2014quantitative,iguchi2017concomitant}.
For all the techniques, it is necessary to take into account that the presence of gas around the thermometer in contact with the thermal bath can limit the precision of the temperature measurement; in order to prevent this condition, it is possible to limit the source of noise that is coming from the gas by keeping the system under vacuum.
The vast majority of experiments are performed with a direct temperature difference measurement by means of sensors such as thermocouples, represented in Fig \ref{fig_1}. This is the most versatile technique because it allows the characterization of thin films on substrates as well as bulk materials or multilayers.
Being this experimental procedure the most used, the round robin comparison reported in the present work is devoted to a determination of the coefficient $S_{\textrm{LSSE}}$ based on the temperature difference measurement.

\IEEEpubidadjcol

\section{The round robin test}

The measurement of the spin Seebeck coefficient $S_{\textrm{LSSE}}$ was carried out in order to investigate its reproducibility in the framework of a collaboration between five institutions: INRiM, Tohoku University, Bielefeld University, Ohio State University and Argonne National Laboratories. The five groups performed the round robin test according to the measurement method described in Fig. 1, i.e. the direct measurement of the temperature difference between the hot and the cold baths that clamp the LSSE sample. This is a single device for all the tests and it is a 4 $\mu$m-thick YIG film grown on a 0.5 mm-thick gadolinium gallium garnet (GGG) substrate, fabricated in Tohoku University. The sample dimensions are 2 mm $\times$ 6 mm and the thickness of the Pt film on the top of the YIG is 10 nm.
The thermal contacts represent the connection between the sample and the thermal baths. In principle, their thermal resistance must be negligible and, regarding the contact over the Pt film, it has to be an electrical insulator in order to avoid the shunt of the Pt layer and, consequently, an interference on the $V_{\textrm{ISHE}}$ measurement. The quality of the thermal contact can improve under the viewpoint of the uniformity by using thermal greases and systems equipped with screws to tighten the sample between the thermal baths. The experimental features adopted by each institution are summarized in table \ref{table1} and described in details in the following sections.

\begin{table*}[!ht]
	\renewcommand{\arraystretch}{1.3}
	\caption{Experimental variables of the electric and thermal measurements of LSSE}
	\label{table1}
	\centering
	\begin{tabular}{|p{2.8cm}||p{3cm}||p{1.5cm}|p{2.2cm}|p{0.5cm}|p{3.1cm}|p{1.5cm}|}			
		\hline
		Institution & Electrical connection & Sensor & Thermal conductor & Area & Thermal grease &$\kappa_{\textrm{grease}}$  \\
		\hline
		&  &  &  & mm$^{2}$ &  & $\textrm{W}\textrm{K}^{-1}\textrm{m}^{-1}$ \\
		\hline
		INRiM & Pt-wires and silver paste & T-type T.C. & AlN & 10 & Rs silicone heat sink paste & 3.6 \\
		\hline
		Tohoku University & Tungsten tips & T-type T.C. & AlN & 10 & Chemtronics Boron Nitride & 1.85 \\
		\hline
		Bielefeld University & Bonding Al-wires & K-type T.C. & Al$_{2}$O$_{3}$ & 10 & Titanium dioxide paste &  0.82 \\
		\hline
		Ohio State University & Cu-wires and silver paste & Cernox & c-BN & 8.76 & Apiezon N & 0.194 \\
		\hline
		Argonne National Labs. & Al-wires and silver paste & K-type T.C. & OFHC-Cu &  8 & Wakefield 122 silicone heat sink paste & 2.5 \\
		\hline
	\end{tabular}
\end{table*}

\subsection{INRiM set-up}
\begin{figure}[h]
	\centering
	\includegraphics[width=3in]{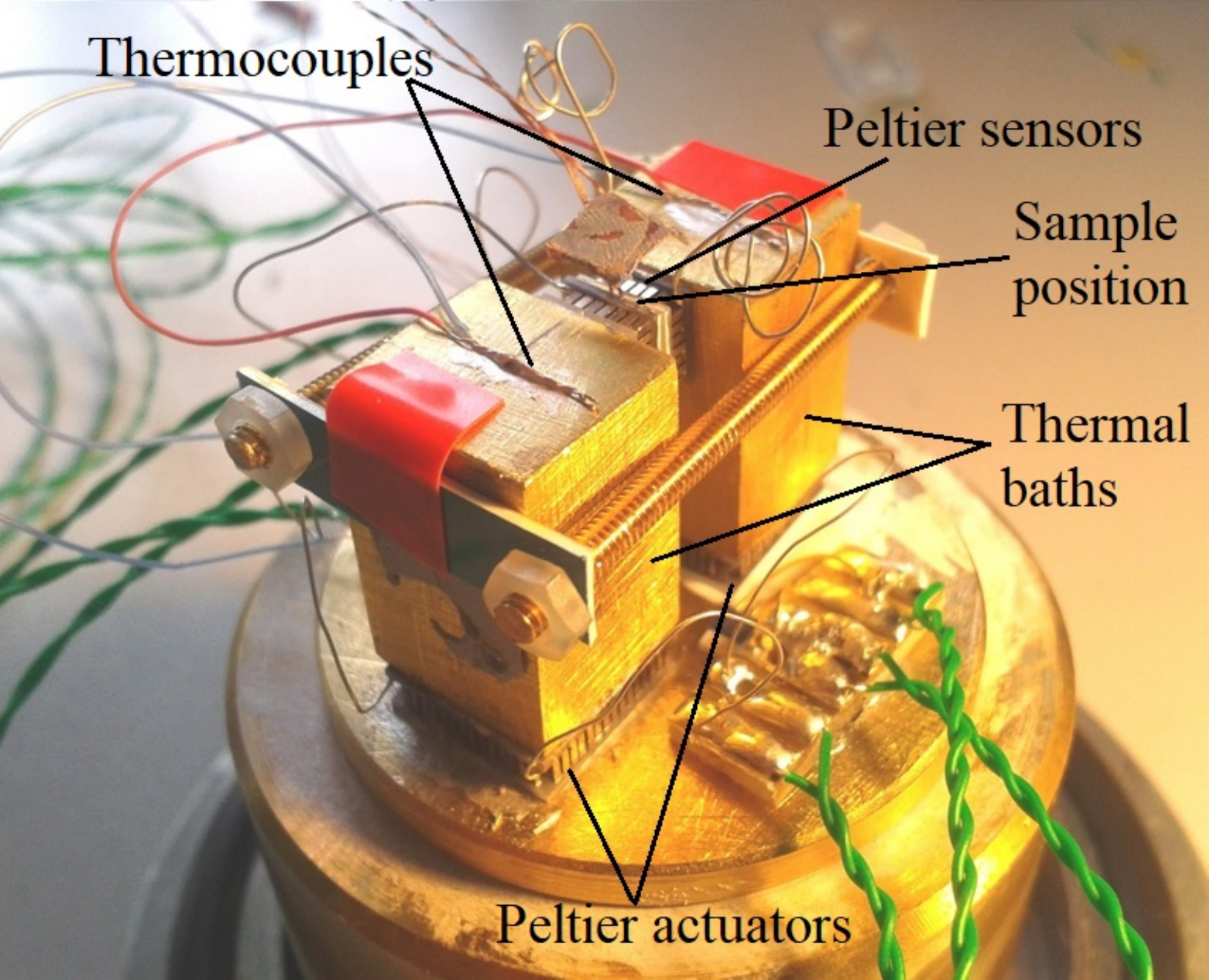}
	\caption{The INRiM measurement system: the electrical contacts are made with Pt wires bonded with silver paste, the temperature sensors are two T-type thermocouples, the thermal conductors are blocks of AIN connected to the LSSE sample with silicone thermal grease.}
	\label{fig_inrim}
\end{figure}

The INRiM measurement system for the LSSE is represented in Fig. \ref{fig_inrim}.

\noindent The temperature gradient is provided by two Peltier elements placed at the bottom of two brass blocks that work as thermal baths whose temperature is monitored by two T-type thermocouples. The system is equipped with two calibrated sensors (Peltier elements) of known thermal conductivity for the measurement of the heat current across the sample under test. The thermal connection on the top of the sample is guaranteed by a 3 mm thick aluminium nitride (AlN) connector with nominal thermal conductivity equal to 140 Wm$^{-1}$K$^{-1}$. The extra temperature drop due to the thermal connector is less that 0.05 K for the quantities of heat involved in this experiment. Each surface of the sample is covered with a thin layer of thermal grease in order to improve the geometrical uniformity of the thermal connection.
The temperature drop due to the heat current sensors is taken into account being their thermal resistance equal to 70 K/W.
The electrical connection on the sample is made with two Pt wires whose diameter is 150 $\mu$m glued with silver paste at the sample edges. The sample is clamped tightly by means of screws and the whole system is under vacuum during the measurements.

\subsection{Tohoku University set-up}

\begin{figure}[h]
	\centering
	\includegraphics[width=3in]{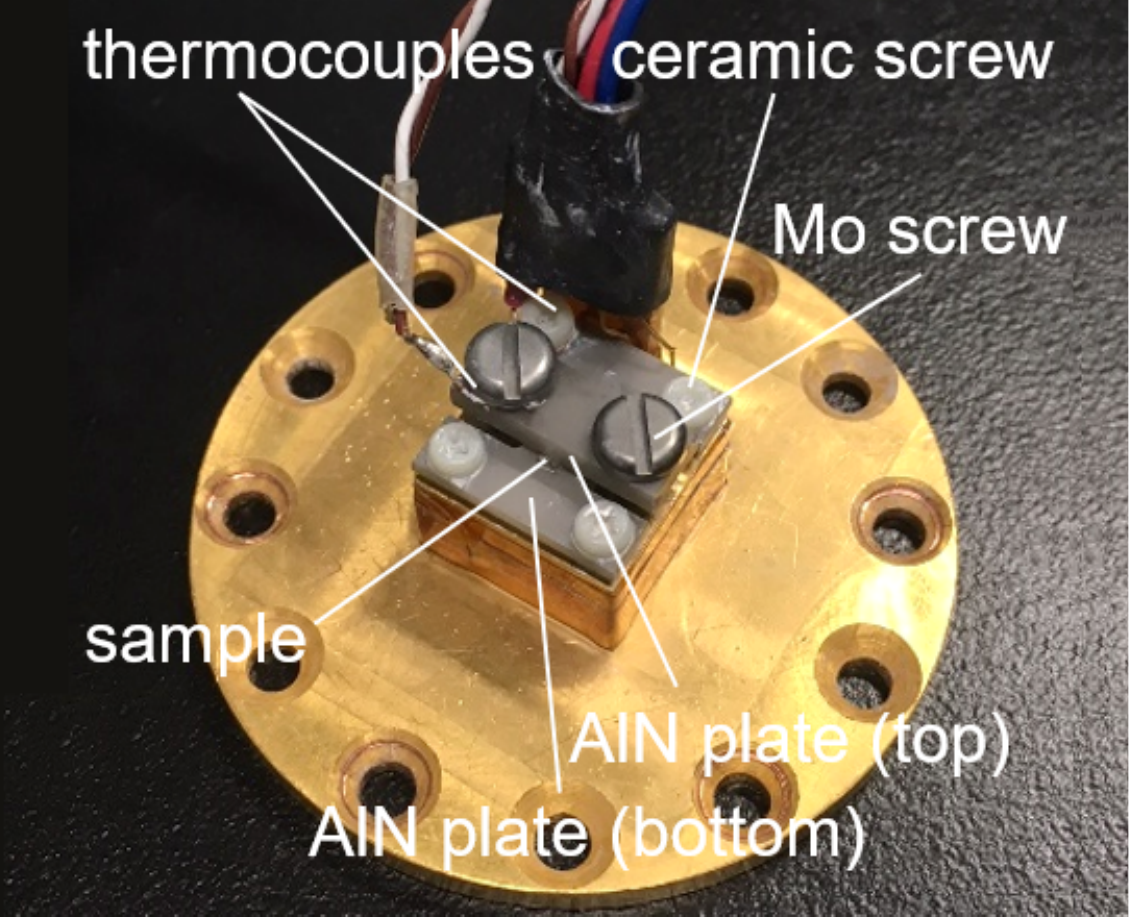}
	\caption{The Tohoku University measurement system: the electrical contacts are made with tungsten needles, the temperature sensors are two T-type thermocouples, the thermal conductors are blocks of AIN connected to the LSSE sample with boron nitride thermal grease. The measurement system has almost the same structure as that shown in Ref. \cite{uchida2012thermal}, but the Cu plates in Ref. \cite{uchida2012thermal} are replaced with the insulating AlN plates.}
	\label{fig_tohoku}
\end{figure}

The Tohoku University measurement system for the LSSE is represented in Fig. \ref{fig_tohoku}.

\noindent The LSSE sample is sandwiched between two AlN plates; the bottom plate is positioned on the top surface of a Peltier thermoelectric module used as temperature difference actuator. The bottom surface of the Peltier module is thermally connected to the heat bath and a temperature gradient is generated in the sample along the $z$ direction by increasing or decreasing the temperature of the lower AIN plate by applying an electric current to the Peltier module. Two T-type thermocouples are measuring the temperature difference between the upper and lower AIN plates.
The measurement of the voltage $V_{\textrm{ISHE}}$ between the ends of the Pt layer of the LSSE sample is performed by means of tungsten needles; these are attached to the ends of the sample by using a micro-probing system.
This configuration is possible since the length of the sample (6 mm) is slightly longer than the width of the upper AIN plate (5 mm). The thermal link between the AIN plate and the sample is guaranteed by a layer of Boron Nitride Heat Sink Grease, ITW Chemtonics. The measurements are performed by keeping the system under vacuum.

\subsection{Bielefeld University set-up}
\begin{figure}[h]
	\centering
	\includegraphics[width=3in]{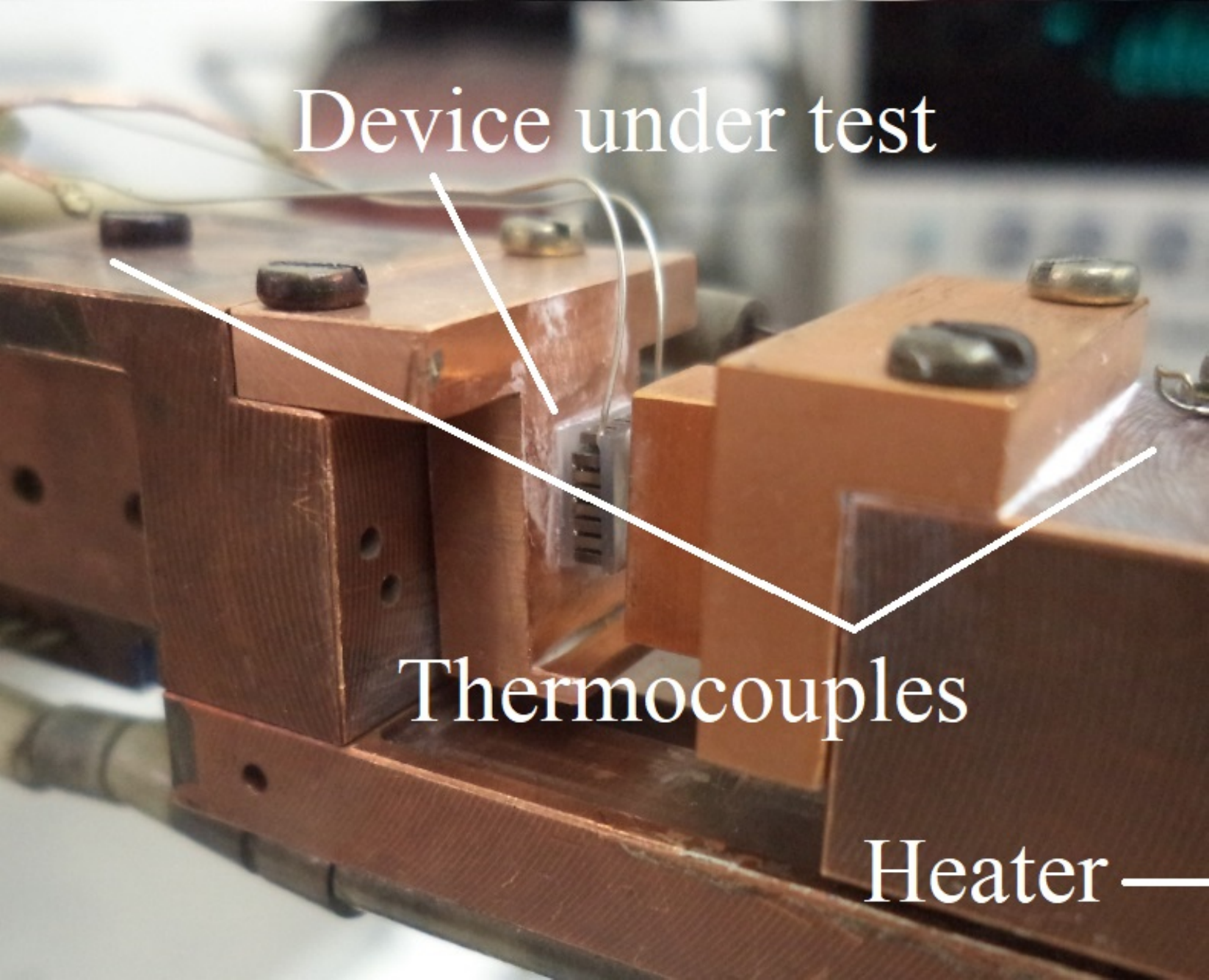}
	\caption{The Bielefeld University measurement system: the electrical contacts are made with bonded Al-wires, the temperature sensors are two K-type thermocouples, the thermal conductors are blocks of AI$_{2}$O$_{3}$ connected to the LSSE sample with titanium dioxide thermal grease.}
	\label{fig_bielefeld}
\end{figure}

The Bielefeld University measurement system for the LSSE is represented in Fig. \ref{fig_bielefeld}.

\noindent The electrical contacts on the top Pt surface of the LSSE sample is obtained by means of a bonding machine that is operating with an Al-wire. The temperature sensors are K-type thermocouples attached to the two thermal baths of which one is heated by means of a resistive heater and the other is connected to a heat sink. Between the copper hot bath and the sample a layer of aluminium oxide (Al$_{2}$O$_{3}$) is positioned in order to guarantee the electrically insulating thermal link. The two surfaces of the LSSE sample under test are covered with a thin layer of titanium dioxide thermal grease that is used in order to improve the uniformity of the thermal connection. The hot bath consists of a copper block mounted on a moving stage that is able to sandwich the sample. The whole system is under vacuum during the characterization of the LSSE sample.
 
\subsection{Ohio State University set-up}
\begin{figure}[h]
	\centering
	\includegraphics[width=3in]{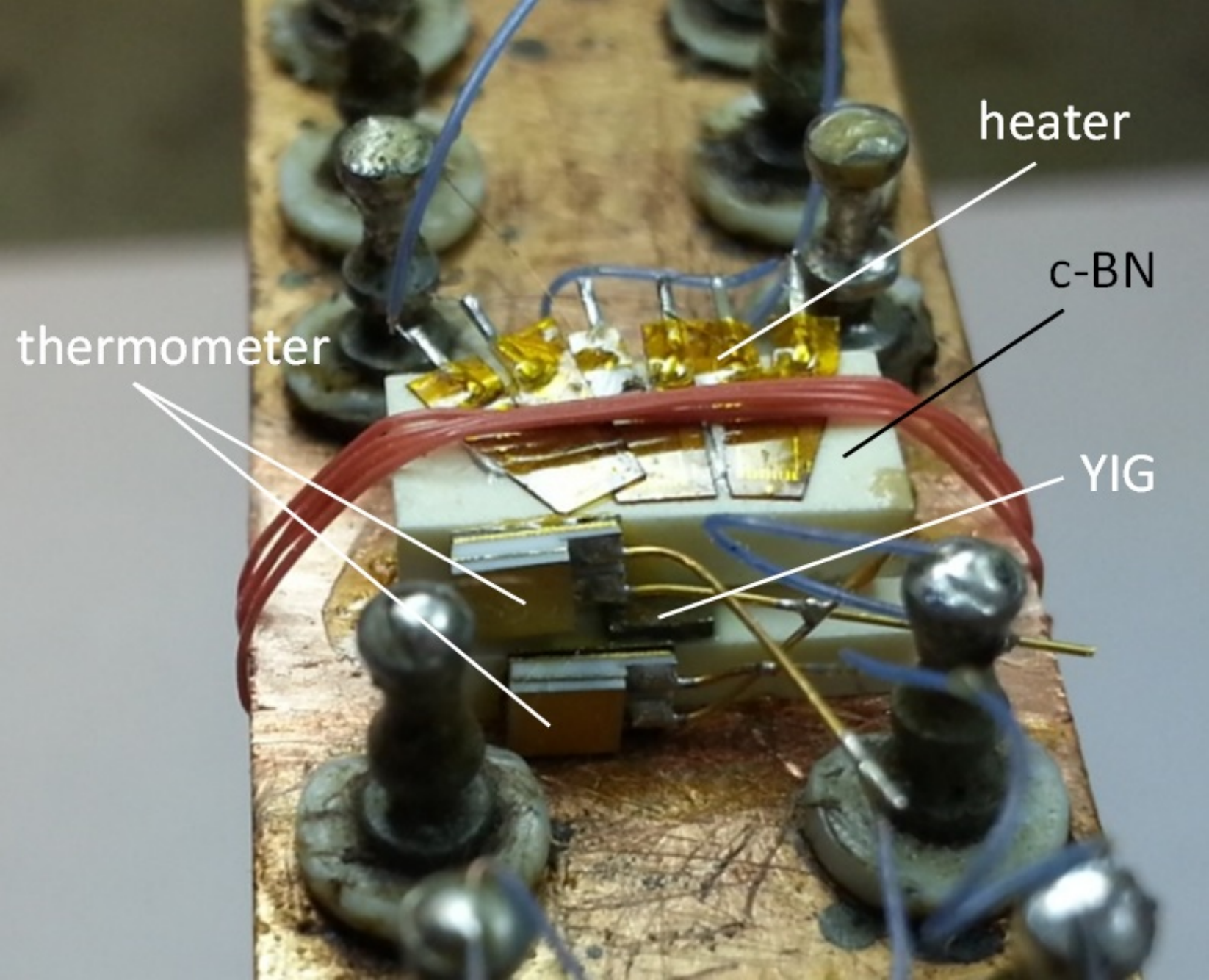}
	\caption{The Ohio State University measurement system: the electrical contacts are made with Cu-wires bonded with silver paste, the temperature sensors are Cernox thermometers, the thermal conductors are blocks of cubic boron nitride connected to the LSSE sample with Apiezon N thermal grease.}
	\label{fig_ohio}
\end{figure}

The Ohio State University measurement system for the LSSE is represented in Fig. \ref{fig_ohio}.

\noindent The electrical contact on the sample is realized with copper wires bonded with Ag-paste. The LSSE sample was sandwiched between two rectangular cubic boron nitride (c-BN) pads. Apiezon N grease was used between the sample and c-BN pads to provide uniform thermal contacts. Three 120 $\Omega$ resistive heaters connected in series were attached on the top of the upper c-BN pad by silver epoxy. A Cernox thermometer was attached to each c-BN pad using Lakeshore VGE-7031 varnish. The whole sample block was then pushed onto the sample platform wherein the Apiezon N grease was used as a thermal contact. The sample block and the platform were wrapped together by an insulated wire, in order to provide a solid thermal contact between them. The whole apparatus is operating under vacuum.

\subsection{Argonne National Laboratories set-up}

\begin{figure}[h]
	\centering
	\includegraphics[width=3in]{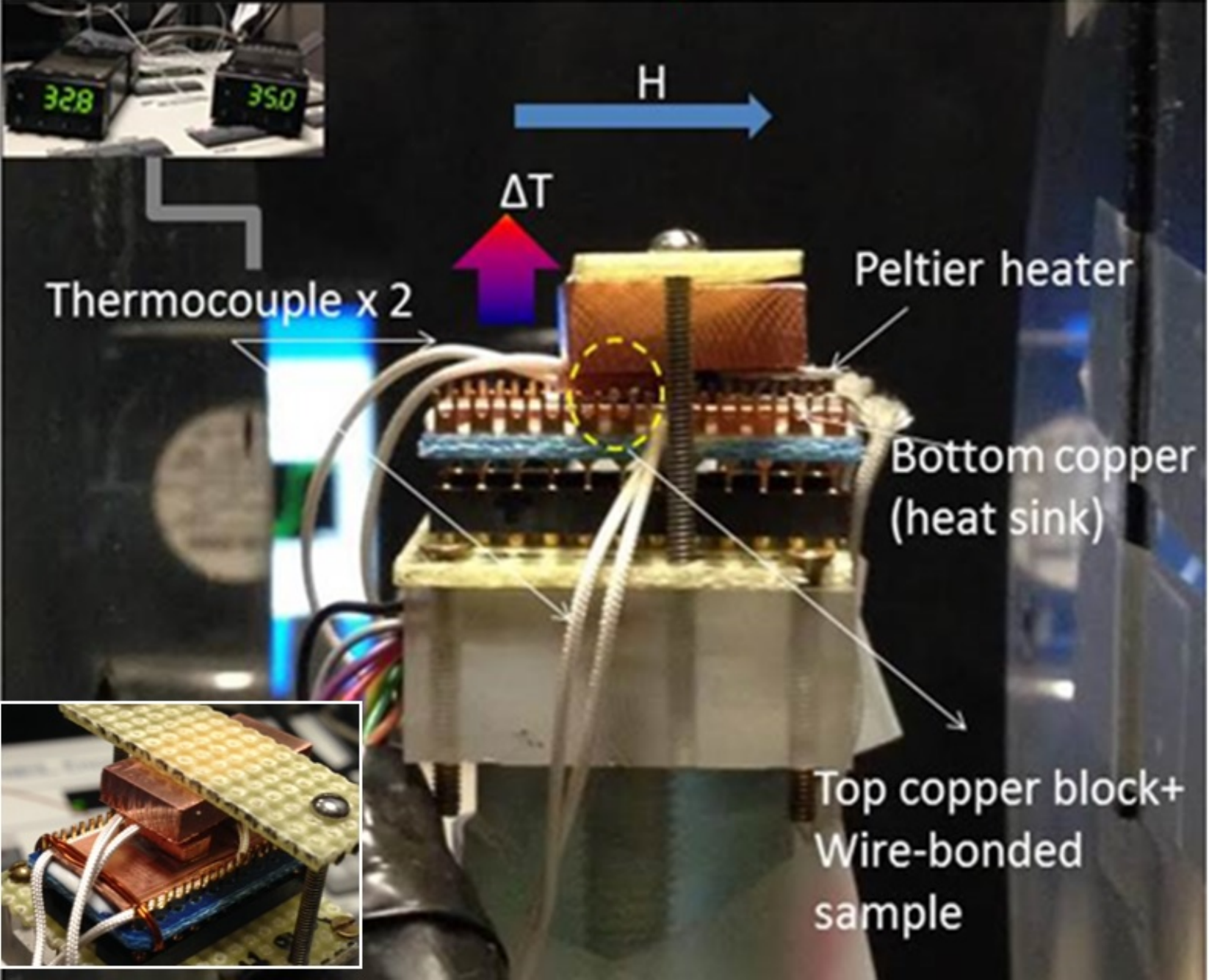}
	\caption{The Argonne National Laboratories measurement system: the electrical contacts are made with Al-wires bonded with silver paste, the temperature sensors are K-type thermocouples, the thermal conductors are blocks of OFHC-Cu connected to the LSSE sample with silicone thermal grease.}
	\label{fig_argonne}
\end{figure}

The Argonne National Laboratories measurement system for the LSSE is represented in Fig. \ref{fig_argonne}. 

\noindent The electrical connection is realized by a plug-pin test holder for convenient loading/unloading samples; two aluminium wires are bonded on the sample with Ag-paste. The electrical holder is topped with a thick Cu pad as heat sink. Two Peltier heaters are used to balance the heat flow: a large (10 mm x 10 mm) Peltier is at the back for cooling the bottom heat sink, and a small (5 mm x 5 mm) Peltier is used to heat the sample. A 5 mm x 5 mm small Cu block is glued with the Peltier for contacting with the sample. Thermal paste is used wherever heat flow exists, i.e., among Cu blocks and between the sample. The extra, up-flowing heat generated from the large Peltier is dissipated through the thick Cu block at the top. Temperature is monitored by two pairs of thermocouple at the top and bottom Cu structures, respectively. The whole apparatus sits on an automated rotating sample stage allowing different magnetic field orientations. The system is under vacuum during the measurements.

\subsection{Results}

The measurement of the voltage $V_{\textrm{ISHE}}$ has as its only variable the ohmic contact between the wire and the film. This contact can be obtained by means of a bonding machine (Bielefeld University), of tungsten tips (Tohoku University) or with the use of silver paste (all other institutions). 
The second observable is the temperature gradient whose measurement is obtained by means of two sensors placed on the thermal baths that are clamping the LSSE sample. The thermal measurement has to fulfill the following hypothesis: the temperature drop along the thermal conductor has to be negligible, the temperature gradient inside the sample is supposed to be constant and the thermal resistance of the contact has to be reproducible.
We assume that all the temperature sensors (T-type, K-type thermocouples and Cernox thermometer) are equivalent as well as the negligible temperature drop across the thermal conductors.
The only variable that can affect the $\Delta T$ measurement is the thermal resistance of the contacts which depends on the quantity, the uniformity and the thermal conductivity of the grease and on the pressure exerted by the screws that clamp the sample.

\begin{figure}[h]
	\centering
	\includegraphics[width=3.5in]{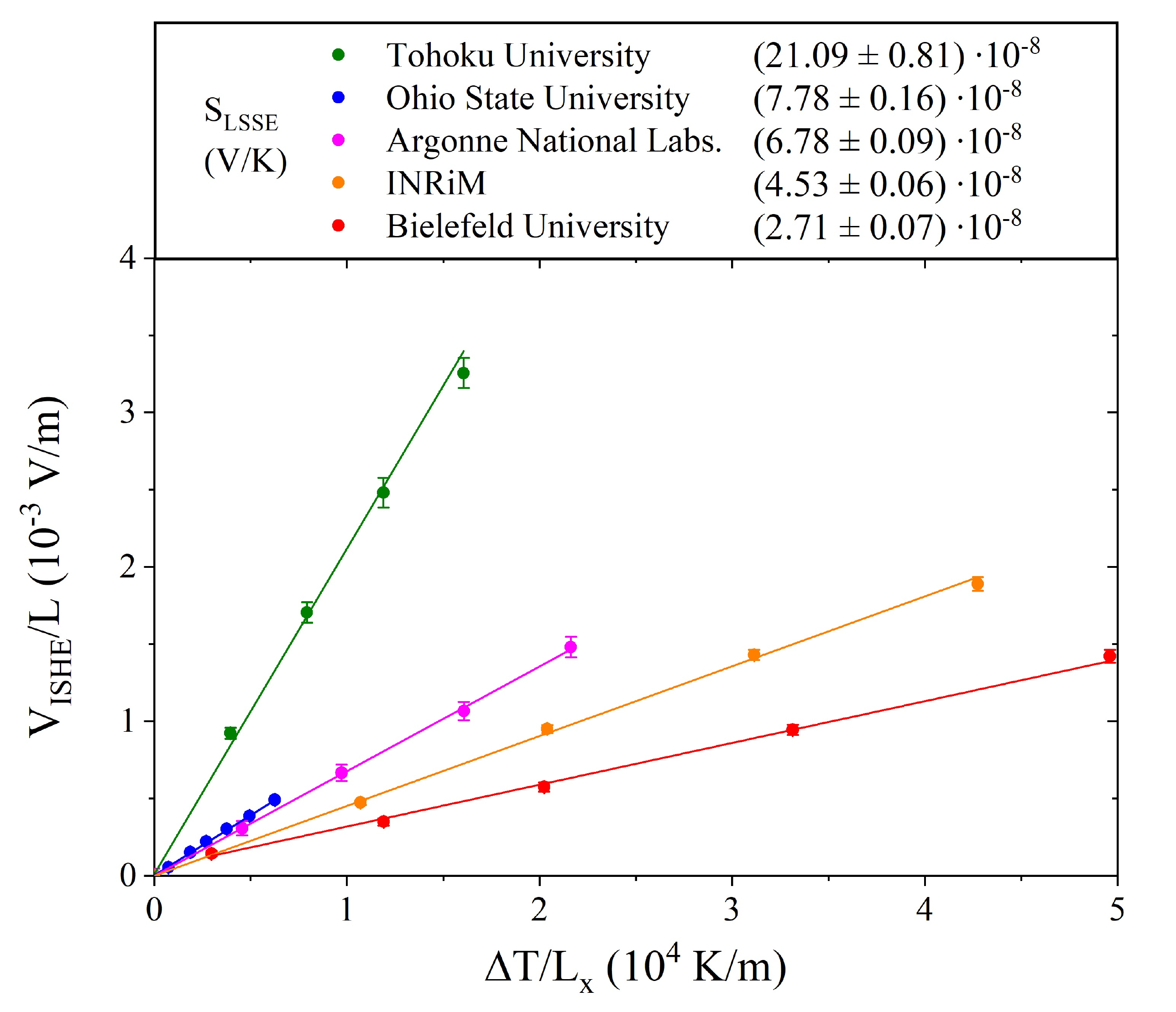}
	\caption{Spin Seebeck coefficients $S_{\textrm{LSSE}}$ obtained by the five partners of the round robin test on a single LSSE device.}
	\label{fig_2}
\end{figure}

The results of this round robin experiment in V/K units exhibit a large variation as reported in Fig. \ref{fig_2}.

\noindent Each point in Fig. \ref{fig_2} is derived from a measurement of the voltage $V_{\textrm{ISHE}}$ at a given temperature difference imposed across the sample. Then, by taking into account the geometrical features of the sample, we can represent the data in terms of electric potential gradient as function of the temperature gradient. The values reported in the label on the top of Fig. \ref{fig_2} are the $S_{\textrm{LSSE}}$ coefficients i.e. the slopes of the linear fits performed by each group; the uncertainties reported here are their errors.
Each group repeats its own measurements so that each point in Fig. \ref{fig_2} is an average value of voltage and the vertical error bar is obtained by the standard deviation. The horizontal error bar is negligible instead, because the temperature difference is set. This points out that the measurement is reproducible internally to each group but the results obtained among the five groups differ by a factor of 8.
We consider the cause of this difference to be a significant presence of systematic effects since each single measurement is reproducible.
In order to investigate the origin of the systematic effects, we perform the analysis of the electric measurement ($V_{\textrm{ISHE}}$) and the thermal one ($\Delta T$), separately.
This requires a different measurement system that allows to keep constant the experimental conditions for one physical observables while perturbing the ones of the other observable. First we keep the thermal measurement constant while changing the electric contacts, then we use the same electric contacts to perturb the conditions for the thermal measurements. This analysis is reported in the following section.


\section{Discussion: analysis of systematic effects}
\subsection{Electric measurement}
\begin{figure}[h]
	\centering
	\includegraphics[width=3.5in]{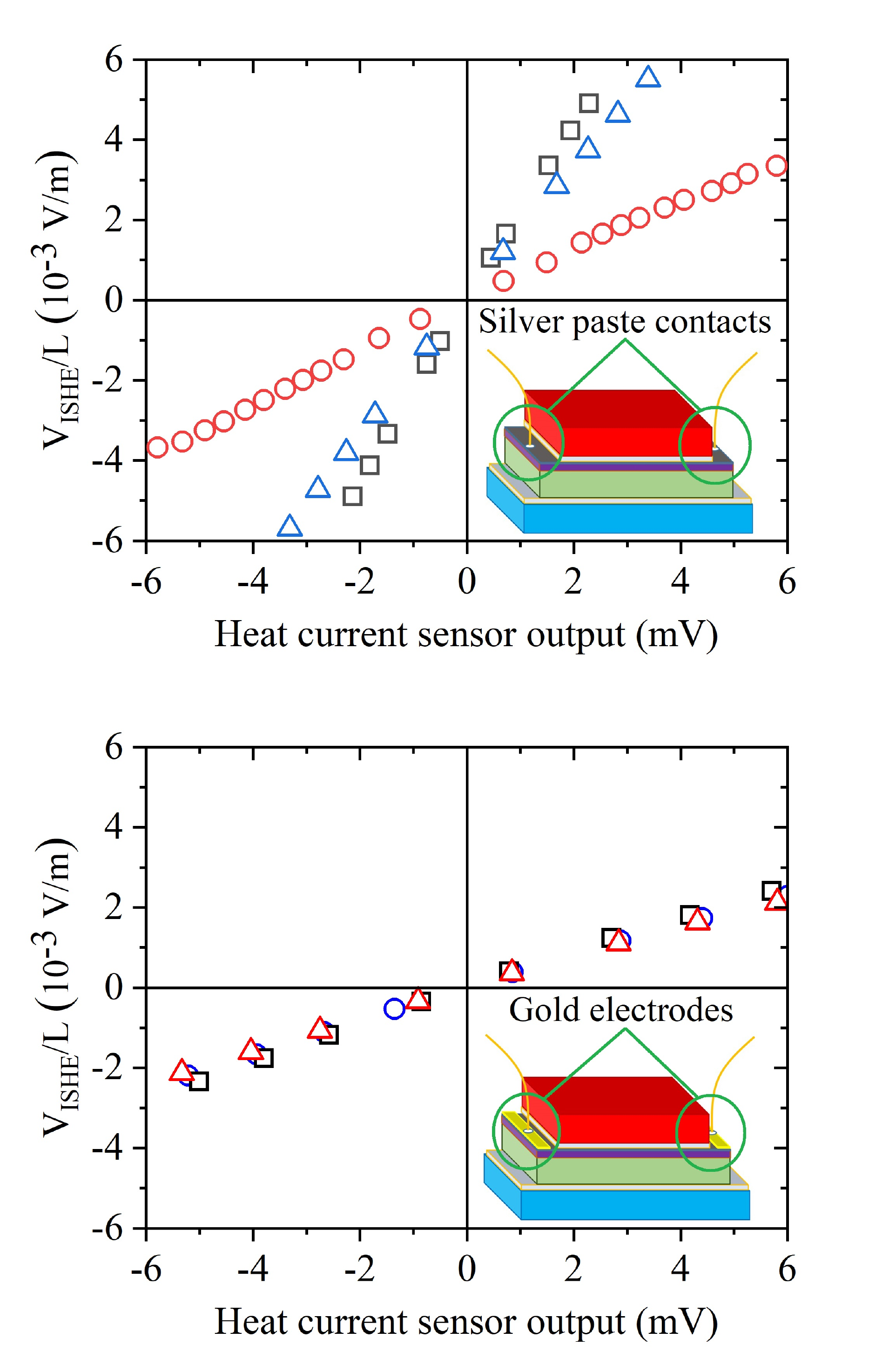}
	\caption{Three data sets of LSSE electric potential gradient as function of the heat current across the sample. The electrical contact is formed with (a) Ag-paste on the bare Pt surface and (b) Ag-paste on the gold electrodes deposited over the Pt layer.}
	\label{fig_3}
\end{figure}

For the analysis of the contribution of the electrical observable $V_{\textrm{ISHE}}$ to the systematic effects, we analysed how the characteristics of the electrical contact influence the $S_{\textrm{LSSE}}$ by performing a separate experiment. The sample as well as the measurement system are reported in this previous work \cite{Sola2} and are different to the ones of the round robin test reported in the previous section. The LSSE sample for this specific test was fabricated at Walther Meissner-Institut, Garching, Germany and it is a 60 nm thick YIG film deposited on 0.5 mm thick yttrium aluminium garnet (Y$_{3}$Al$_{5}$O$_{12}$) single crystal substrates with 5 $\times$ 2.5 mm$^{2}$ in dimension.
The thermal variables have to be fixed by hypothesis, since this experiment is devoted to the study of the ohmic contact as the only source of errors. In order to guarantee this condition, we keep the same device mounting in a single measurement system for the two groups of measurements. This specific system is probing the heat current across the sample instead of the temperature difference between the hot and the cold baths.
For the first group, we changed the shape of the electrical contacts for each series of measurements by changing the size of the silver paste contacts. For the second group, we set a well-defined shape of the electric contacts by depositing gold electrodes at the edges of the Pt surface and we repeated the previous series of measurements.
The results of this side-experiment are reported in Fig. \ref{fig_3}.

From Fig. \ref{fig_3} (a) it's evident that the uncontrolled shape of the electrical contacts on the bare surface of the Pt layer causes a lack of reproducibility between the three sets of data. This results in an error that can be minimized thanks to the controlled geometry of electrical contacts (Fig. \ref{fig_3} (b) experimental configuration represented in the inset).
Although this supplementary experiment has been carried out with a different sample and measurement system with respect to the round robin experiment, it is possible to perform a comparison between the variation found in Fig. \ref{fig_3} (a) with the one reported in Fig. \ref{fig_2}. Assuming a rectangular distribution for the values of the slopes of the two data sets, the relative uncertainty is 32\% for the data reported in Fig. \ref{fig_3} (a), while concerning the round robin results the relative uncertainty is 45\%. This could mean that the electric variable has been partially perturbed by changing the size of the silver paste contacts, in comparison to the scenario of the round robin experiment. Another interpretation is that the electrical observable is not the only source of systematic effects, as discussed in the following section. 

\subsection{Thermal measurement}
Another source of errors concerns with the thermal resistance of the contacts between the thermal conductors and the sample under test, since a correct evaluation of the $\Delta T$ across the active magnetic layer is possible only if the exact value of the thermal contact resistance is known.
Thermal contact area as well as thermal conductivity of the contact on the top of the Pt film (fourth column of Table \ref{table1}) condition the $S_{\textrm{LSSE}}$ coefficient. However it's clear from the results plotted in Fig. \ref{fig_2} that the contact area is not the major source of systematic effects, since the results of the two groups that are connecting the same area (Bielefeld University and Tohoku University) are the ones that differ most.
It is reasonable to consider as a crucial feature the quantity and the uniformity of the thermal grease together with the pressure exerted by means of the screws, which influence strongly the thermal conductivity of the contact. Indeed, contextually to the round robin test, it was shown that the different conditions of contacts yield to variations of their thermal resistance as large as the one of the LSSE sample itself \cite{Sola1}.

The idea of using two Pt films deposited on the two faces of the sample for the measurements of the temperature gradients is in favour of solving the problem of the thermal resistance of the contacts \cite{uchida2014quantitative,iguchi2017concomitant}.
In view of the solution of this problem, a second method was proposed; this is based on the measurement of heat currents and allows to neglect the contribution of the thermal resistance of the contacts on a sample with known thermal conductivity.
The accuracy of this method depends on the assumption that all the elements of the thermal circuit (sample and sensors) are in series, i.e. with negligible heat leakage along the circuit.
A model of this method is represented by an electrical analogue of a current measurement in a circuit with unknown resistors in series; these are the thermal contacts at the interfaces between the sample and the heat baths.

In order to validate this method, a second comparison was performed between INRiM and Bielefeld University.
This test is also an additional experiment and it has been performed with the sample described in the previous section, which is not the one of the round robin experiment.
For this second comparison, the sample has been equipped with gold electrodes in order to freeze the electric observable and investigate the source of errors originating from the thermal one. This corresponds to maintain the reproducibility achieved in the previous test whose result is reported in Fig. \ref{fig_3} (b). Therefore, it is possible to analyse the thermal observable by comparing results obtained by two measurement techniques: the one used for the round robin experiment, i.e. the measurement of temperature differences and the new method, based on the detection of heat currents.
Each group prepared a measurement system based on the each method and, from the characterisation of a single LSSE device, the $S_{\textrm{LSSE}}$ coefficients obtained by the two institutions was $(9.716\pm 0.060)\cdot 10^{-7}$ V/K and $(9.359\pm 0.128)\cdot 10^{-7}$ V/K with the heat current measurement method and $(2.313\pm 0.017)\cdot 10^{-7}$ V/K and $(4.956\pm 0.005)\cdot 10^{-7}$ V/K with the temperature difference method \cite{Sola2}.
This result points out the advantage of controlling the heat currents and fulfill the hypothesis of having negligible heat leakages rather that having reproducible values of thermal resistances between the sensors and the sample. However, in order to compare the results obtained with the heat current method with the ones determined by the temperature gradient technique it is necessary the information about its thermal conductivity, which itself involves a measurement of temperature difference.
Beyond the framework of the LSSE characterizations, this comparison highlights a more general issue of the measurements of temperature gradients across thin films\cite{zhang2015thermal,huebner2018thermal}; such a configuration make the handling of the thermal resistances of the contacts particularly adverse and favoured the development of other measurement techniques for the spin caloritronics.

\section{Conclusion}
This work is the extended version of the digest presented at the CPEM 2018 (Conference on Precision Electromagnetic Measurements) \cite{sola2018cpem}.
A round robin test for the measurement of the spin Seebeck coefficient $S_{\textrm{LSSE}}$ pointed out the experimental problems for its quantitative determination. This was carried out with the characterisation of a single LSSE device by five different research groups; each of them developed a measurement system based on the detection of temperature differences. The results obtained range from $2.7\cdot 10^{-8}$ V/K to $21.1\cdot 10^{-8}$ V/K and are therefore affected by systematic effects.
The contribution from the systematic effects to the coefficient $S_{\textrm{LSSE}}$ was discussed, both for the electrical observable $V_{\textrm{ISHE}}$ and the thermal one ($\Delta T$).
For what concerns the electrical observable $V_{\textrm{ISHE}}$, the deposition of gold electrodes at the edges of the Pt film has been proved to be a viable solution.
About the $\Delta T$ measurement, the role of the thermal resistance of the contact is crucial in terms of source of systematic effects.
This was proven by a separate comparison in which the $S_{\textrm{LSSE}}$ coefficient is obtained by two different methods that are based on the temperature difference and the heat current measurements; the last method does not depend on the thermal resistance of the contact and it allows more reproducible measurements.
In general, the search for a method that allows reproducible measurement is crucial for the quantitative characterization of different LSSE materials and other spin caloritronics effects such as the spin Peltier effect.


%



\section*{Acknowledgment}

This work has been carried out within the Joint Research Project EXL04 (SpinCal) and the Deutsche Forschungsgemeinschaft (DFG) within the priority program Spin Caloric Transport (SPP1538, KU 3271/1-1, RE 1052/24-2).
Measurements at Argonne were supported by the US Department of Energy, Office of Science, Materials Science and Engineering Division.
The group from INRiM thanks Prof. Dr. Mathias Kl\"{a}ui for fruitful discussion and sample preparation.

\ifCLASSOPTIONcaptionsoff
  \newpage
\fi

\end{document}